\documentclass[12pt,a4paper,final]{iopart}

\usepackage{iopams}  
\usepackage[breaklinks=true,colorlinks=true,linkcolor=blue,urlcolor=blue,citecolor=blue]{hyperref}
\usepackage{enumitem}
\usepackage{graphicx}
\usepackage{pifont}
\newcommand{\cmark}{\ding{51}}%
\newcommand{\xmark}{\ding{55}}%

\begin{document}

\title[Phase separation can be stronger than chaos]{Phase separation can be stronger than chaos}

\author{Andrea Richaud and Vittorio Penna}
\address{ Dipartimento di Scienza Applicata e Tecnologia and u.d.r. CNISM, Politecnico di Torino, 
Corso Duca degli Abruzzi 24, I-10129 Torino, Italy}
\ead{andrea.richaud@polito.it}

\begin{abstract}
We investigate several dynamical regimes characterizing a bosonic binary mixture loaded in a ring trimer, with particular reference to the persistence of demixing. The degree of phase separation is evaluated by means of the ``Entropy of mixing", an indicator borrowed from Statistical Thermodynamics. Three classes of demixed stationary configurations are identified and their energetic and linear stability carefully analyzed. An extended set of trajectories originating in the vicinity of fixed points are explicitly simulated and chaos is shown to arise according to three different mechanisms. In many dynamical regimes, we show that chaos is not able to disrupt the order imposed by phase separation, i.e. boson populations, despite evolving in a chaotic fashion, do not mix. This circumstance can be explained either with energetic considerations or in terms of dynamical restrictions.

\end{abstract}


\section{Introduction} 
The demixing of the condensed species constituting a binary bosonic mixture, i.e. their localization in different spatial regions, is a process that can be triggered by the presence of strong inter-species repulsive interactions. The underlying mechanism of phase separation (also called species separation) has been thoroughly investigated within the field of ultracold bosons by means of the mean-field representation of condensate dynamics \cite{cmixt3,cmixt4,cmixt5}. The onset of the demixing transition has been shown to depend on a number of factors, including the shape of the trapping potentials, the number of bosons in each atomic species and the interaction parameters. At the same time, considerable attention has been devoted to the dynamical-stability analysis and to the appearance of excited states in regimes close to the transition point. \cite{kasa,Gallemi_1,tic,lee}. 

Engineering of experimental setups capable of trapping boson mixtures \cite{Inguscio,Gadway,Soltan} in optical lattices \cite{jz,bdz,yuk} has stimulated the interest of the theoretical community for the understanding of demixing effect in the presence of spatial fragmentation. In this context, beyond phase separation \cite{sep2,sep3, Angom}, a rich variety of phenomena have been highlighted including (but not limited to) the formation of boson currents in ring lattices, \cite{NoiPRA2}, the appearance of novel phases exhibiting magnetic-like properties \cite{ks,ddl}, quantum emulsions \cite{qe1}, the emergence of polaron excitations \cite{pol,Tempere}, the entanglement between the species \cite{ent, NoiEntropy}, the collision of a condensate with impurities \cite{Penna_NJP,Catani,EPL}, the formation of immiscible solitons \cite{Makarov} and the modulation instability in the phase separation \cite{modul}.
Systems consisting of two bosonic species and optical lattices with ring geometry are within the reach of current experimental setups. In particular, heteronuclear and homonuclear mixtures have been realized in \cite{Inguscio, Catani_deg} and \cite{Gadway, Soltan_2} respectively, while the ring-lattice geometry has been designed and employed by \cite{Amico, Agamalian}. Moreover, population oscillations can be monitored with the same techniques used to reveal the self-trapping effect of a single condensate in a two-well system \cite{Anker, Albiez}.

Recently, the phase separation mechanism of a binary bosonic mixture has been investigated in the two simplest but non-trivial lattice geometries: the double well \cite{NoiEntropy,PennaLinguaJPB,PennaLinguaPRE} and the ring trimer \cite{NoiSREP}. It has been evidenced that in the former case only one kind of transition occurs, while in the latter case the demixing develops in two steps, i.e. an intermediate neither fully mixed nor completely demixed phase exists. Moreover, it has been shown that in the parameters' space of the model, the critical point where phase separation occurs is characterized by the collapse and rearrangement of the energy levels and by singularities in the entanglement entropy between the two species \cite{NoiEntropy,NoiSREP}.

Provided that the number of bosons is large, one can consider the semiclassical counterpart of the quantum system and investigate the dynamics generated by the resulting set of discrete nonlinear Schr\"{o}dinger equations \cite{Penna_Amico, Nonlinearity, Eilbeck}. The latter, which constitute the discrete analogue of the Gross-Pitaevskii equation, feature an extraordinary rich scenario of dynamical regimes. Already in the relatively simple case of a single condensed species loaded in a ring trimer, one can identify both stable and unstable regimes including vortices, dimerlike states and chaotic oscillations \cite{Penna_PRE,Johansson,Olsen}. In \cite{Arwas}, the non-integrable character of low dimensional circuits has been evidenced and chaos has been shown to support the persistence of superfluidity.   

In this work we investigate, by means of a semiclassical approach, the dynamics of a bosonic binary mixture loaded in a ring trimer, emphasizing its relation with the entropy of mixing and the persistence of spatial phase separation. After identifying three classes of stationary configurations featuring an high degree of demixing and after developing the energetic- and the linear-stability analysis, we simulate the dynamics of thousands of trajectories starting in the vicinity of fixed points (FPs). These simulations (i.e. the numerical solutions of motion equations (\ref{eq:z_dot})) not only allow one to compute the first Lyapunov exponent, an indicator which allows to distinguish between regular and chaotic trajectories, but also give the possibility to monitor the degree of mixing of the two condensed species. Contrary to expectations, we show that there are several dynamical regimes where chaos, despite present, is not able to disrupt the order imposed by spatial phase separation. This circumstance will be clarified both with dynamical considerations and in terms of energy conservation.

The outline of the manuscript is as follows: in section \ref{sec:A_binary_mixture} we present the model and its semiclassical counterpart. In section \ref{sec:Notable_classes}, we identify three notable classes of stationary configurations featuring spatial phase separation. Section \ref{sec:Stability_of} is devoted to the analysis of their energetic and linear stability. In section \ref{sec:Regular_and_chaotic}, we perform our numerical simulations and we compute the first Lyapunov exponent. An indicator to quantify the degree of mixing of the atomic species is presented in section \ref{sec:Quantify_mixing}. In section \ref{sec:Despite_chaos}, we discuss some meaningful dynamical regimes, putting particular emphasis on those ones where chaos and persistent demixing coexist. Eventually, section \ref{sec:Conclusions} is devoted to concluding remarks.

\section{A binary mixture in a ring trimer}
\label{sec:A_binary_mixture}
The second-quantized Hamiltonian describing a bosonic mixture of two atomic species in a three-well potential (with periodic boundary conditions) is 
$$
  \hat{H}= - T_a \sum_{j=1}^{3} \left(\hat{a}_{j+1}^\dagger \hat{a}_j +\hat{a}_j^\dagger \hat{a}_{j+1} \right) + \frac{U_a}{2} \sum_{j=1}^{3} \hat{n}_j(\hat{n}_j-1)
$$
$$  - T_b \sum_{j=1}^{3} \left(\hat{b}_{j+1}^\dagger \hat{b}_j +\hat{b}_j^\dagger \hat{b}_{j+1} \right)
  +\frac{U_b}{2} \sum_{j=1}^{3} \hat{m}_j(\hat{m}_j-1)
$$
\begin{equation}
\label{eq:Hamiltoniana_quantistica}
  +W \sum_{j=1}^{3} \hat{n}_j\, \hat{m}_j 
\end{equation}  
where $j=4\equiv 1$ due to the ring geometry. This is a typical Bose-Hubbard Hamiltonian, where $T_a$ and $T_b$ are the tunnelling amplitudes, $U_a$ and $U_b$ represent \textit{intra-}species repulsive interactions and $W$ corresponds to the \textit{inter-}species repulsion. Creation and destruction operators satisfy usual bosonic commutators, namely $[\hat{a}_i,\hat{a}_j^\dagger]=[\hat{b}_i,\hat{b}_j^\dagger]=\delta_{i,j}$ and $[\hat{a}_i,\hat{b}_j]=[\hat{a}_i,\hat{b}_j^\dagger]=0$. $\hat{n}_j=\hat{a}_j^\dagger \hat{a}_j$ and $\hat{m}_j=\hat{b}_j^\dagger \hat{b}_j$ are number operators and their sums, $\hat{N}=\sum_{j=1}^3\hat{n}_j$ and $\hat{M}=\sum_{j=1}^3 \hat{m}_j$ respectively, constitute two independent conserved quantities, being $[\hat{N},\hat{H}]=[\hat{M}, \hat{H}]=0$. Provided that the number of bosons is sufficiently high \cite{Kolovsky}, it is possible to replace field operators in Hamiltonian (\ref{eq:Hamiltoniana_quantistica}) with local order parameters, \cite{Penna_Amico, Ellinas}. Such substitutions, which explicitly read
$$
  \hat{a}_j \quad \rightarrow  \quad a_j, \qquad   \hat{b}_j \quad \rightarrow  \quad  b_j,
$$
allow one to cast the quantum dynamics generated by Hamiltonian (\ref{eq:Hamiltoniana_quantistica}) in a classical form, that is
$$
  i\hbar\dot{a}_j = -T_a(a_{j-1}+a_{j+1})+a_j\left(U_a|a_j|^2+W|b_j|^2 \right)
$$
$$
  i\hbar\dot{b}_j = -T_b(b_{j-1}+b_{j+1})+b_j\left(U_b|b_j|^2+W|a_j|^2 \right).
$$
It is convenient to express local order parameters in terms of number of bosons and local phase \cite{Arwas,Kolovsky}, i.e. $a_j =  \sqrt{n_j}e^{i\phi_j}$ and $ b_j=\sqrt{m_j}e^{i\psi_j}$. One thus obtain the following classical Hamiltonian 
$$
 H =-2T_a \left(\sqrt{n_2 n_1} \cos \left(\phi_2- \phi_1  \right)+\sqrt{n_3 n_2} \cos \left(\phi_3- \phi_2 \right)+\sqrt{n_1 n_3} \cos \left(\phi_1- \phi_3  \right) \right)
$$
$$
  -2T_b \left(\sqrt{m_2 m_1} \cos \left(\psi_2- \psi_1  \right)+\sqrt{m_3 m_2} \cos \left(\psi_3- \psi_2 \right)+\sqrt{m_1 m_3} \cos \left(\psi_1- \psi_3  \right) \right)
$$
$$
  + \frac{U_a}{2}\left( n_1^2+n_2^2+n_3^2 \right) + \frac{U_b}{2}\left( m_1^2+m_2^2+m_3^2 \right)
$$ 
\begin{equation}
    \label{eq:Hamiltoniana_classica}
  +W(n_1m_1+n_2m_2+n_3m_3)
\end{equation}
which, in turn, after setting $\hbar=1$, entails the following motion equations 
\begin{equation}
  \dot{\phi}_j = \frac{\partial H}{\partial n_j}, \qquad \dot{n}_j=- \frac{\partial H}{\partial \phi_j},
  \label{eq:Hami_1}
\end{equation}
\begin{equation}
  \dot{\psi}_j = \frac{\partial H}{\partial m_j}, \qquad \dot{m}_j=- \frac{\partial H}{\partial \psi_j}.
   \label{eq:Hami_2}
\end{equation}

\section{Notable demixed stationary configurations}
\label{sec:Notable_classes}
The exhaustive study of all possible stationary configurations (i.e. configurations having a trivial time dependence) which the system admits goes beyond the scope of this work, as the scenario is extraordinarily branched and rich. Already with a single condensed species confined in a ring trimer, several classes of stationary states (e.g. vortex, $\pi$ and dimerlike states) have been evidenced \cite{Penna_PRE}.

\begin{figure}[htb!]
 \centering
 \includegraphics[width=\linewidth]{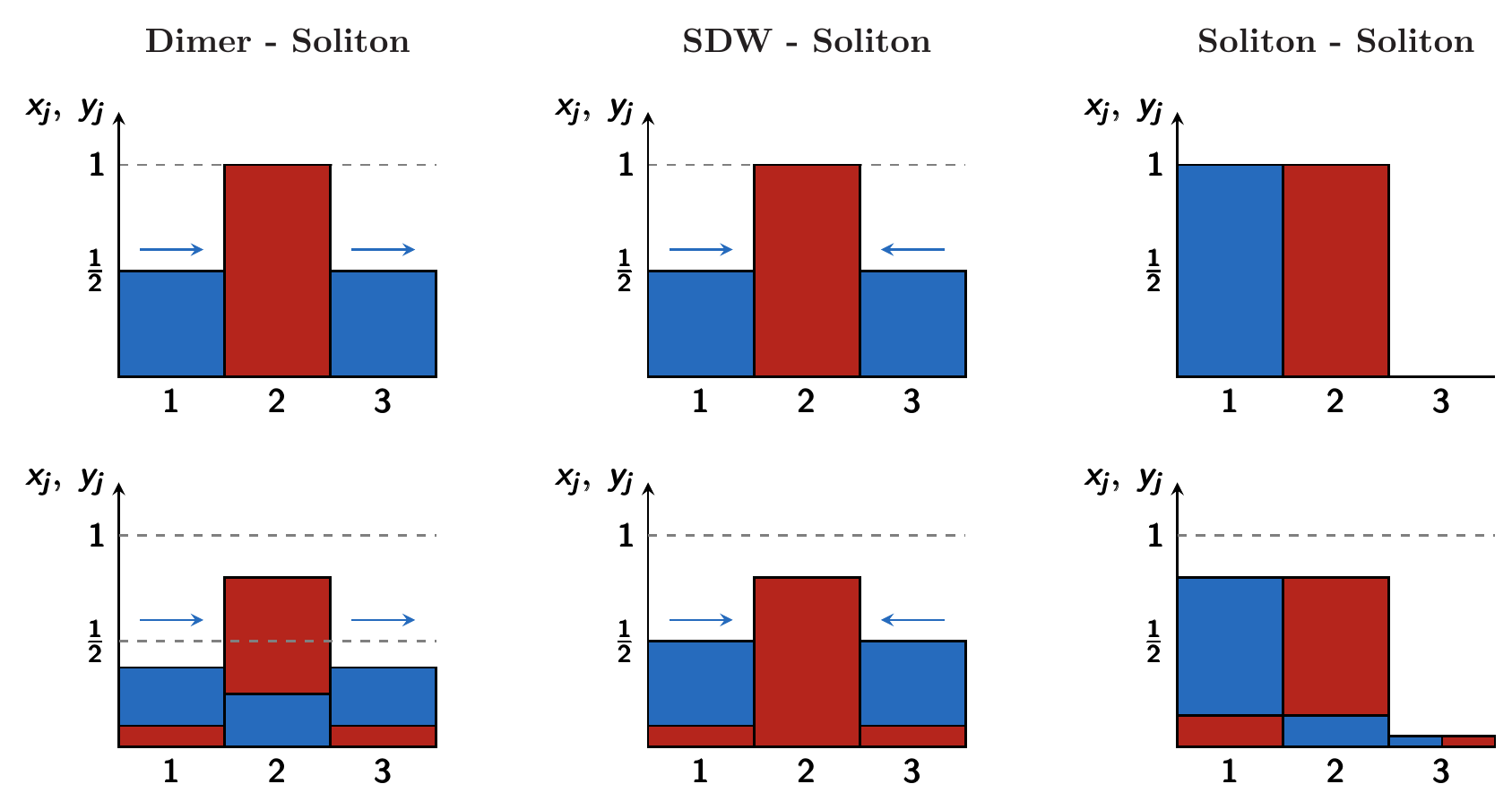}
 \caption{The three families of stationary, demixed, configurations for zero (upper row) and non-zero (lower row) hopping amplitudes $T_a$ and $T_b$. Lower row displays in an exaggerate but illustrative manner the deviations from the zero-tunnelling scenario. Numbers on the horizontal axis correspond to wells' labels, while the height of the histograms represents normalized populations $x_j=n_j/N$ and $y_j=m_j/M$. Parallel (antiparallel) arrows stand for ``in-phase" (antiphase) condensates. Wherever not explicitly defined, condensate phases assume different values according to different choices of model parameters (see \ref{sec:App_Stat_Config}). }
 \label{fig:3_families}
\end{figure}

In this work, we put the focus onto the miscibility properties of the two condensed species, and we start our analysis from some notable stationary configurations which feature phase separation. According to the theory of discrete nonlinear Schr\"{o}dinger equations \cite{Kevrekidis}, substitutions $ \phi_j \to \Phi_j +\lambda_a t $ and $  \psi_j \to \Psi_j + \lambda_b t$, where $\lambda_a$ and $\lambda_b$ represent collective angular frequencies of condensates' phases, constitute a preliminary step in the search for stationary configurations (the presence of two independent collective frequencies $\lambda_a$ and $\lambda_b$, follows from the density-density form of the interspecies interaction of Hamiltonian (\ref{eq:Hamiltoniana_quantistica})). These substitutions, in fact, allow one to recast Hamilton equations (\ref{eq:Hami_1}) and (\ref{eq:Hami_2}) into the following dynamical system
\begin{equation}
\label{eq:Dyn_sys}
\!\!\!\!\!\!\!\!\!\!\!\!\!\!\!\!\!\!\!\!\!\!\!\!\!\!\!\!\!\!\!\!\!\!\!\left \{
\begin{array}{c}
   \dot{\Phi}_j  = U_a n_j + Wm_j-\lambda_a -T_a \left[\sqrt{\frac{n_{j-1}}{n_j}}\cos(\Phi_{j-1}-\Phi_j)   +   \sqrt{\frac{n_{j+1}}{n_j}}\cos(\Phi_{j+1}-\Phi_{j}) \right]\\
   \\
   \dot{\Psi}_j = U_b m_j + Wn_j-\lambda_b-T_b \left[\sqrt{\frac{m_{j-1}}{m_j}}\cos(\Psi_{j-1}-\Psi_j)   +   \sqrt{\frac{m_{j+1}}{m_j}}\cos(\Psi_{j+1}-\Psi_{j}) \right] \\
   \\
  \dot{n}_j=2T_a \left[\sqrt{n_{j-1} n_j}\sin(\Phi_{j-1}-\Phi_j) +\sqrt{n_jn_{j+1}} \sin(\Phi_{j+1}-\Phi_j) \right]\\
  \\
  \dot{m}_j=2T_b \left[\sqrt{m_{j-1} m_j}\sin(\Psi_{j-1}-\Psi_j) +\sqrt{m_jm_{j+1}} \sin(\Psi_{j+1}-\Psi_j) \right]
\end{array}
\right.
\end{equation}
with $j=1,2,3$. Looking for FPs of the latter (which therefore correspond to stationary solutions of equations (\ref{eq:Hami_1}) and (\ref{eq:Hami_2})), together with the two constraints $\sum_{j=1}^3 n_j =N $ and $\sum_{j=1}^3m_j=M$, one finds three classes of configurations which, in the limit $T_a,T_b \to 0$, feature perfect demixing. They are schematically illustrated in figure \ref{fig:3_families} (upper row) and described below: 
\begin{enumerate}
    \item \textit{Dimer - Soliton:} Condensate A is equally subdivided in two wells, the phases therein being the same, while the third well contains all the condensate B.
    $$
       \begin{array}{cccc}
        n_1=N/2,    & n_2=0,  & n_3=N/2, & \lambda_a=NU_a /2,  \\
        m_1=0,      & m_2=M,  & m_3=0,   & \lambda_b= MU_b. 
       \end{array}
    $$
    \item \textit{Single Depleted Well (SDW) - Soliton:} Condensate A is equally subdivided in two wells but, contrary to the previous case, the relative phase $\Phi_3-\Phi_1$ between such wells is $\pi$. The third well contains all the condensate B.
    \item \textit{Soliton - Soliton:} One well contains all the condensate A while an other well contains all the condensate B.
    $$
       \begin{array}{cccc}
        n_1=N,    & n_2=0,  & n_3=0, & \lambda_a=NU_a,  \\
        m_1=0,    & m_2=M,  & m_3=0, & \lambda_b= MU_b. 
       \end{array}
    $$
\end{enumerate}
Upon activation of hopping  amplitudes $T_a$ and $T_b$, FPs slightly deviate from the aforementioned ones, as some bosons move from the macroscopically occupied wells to neighbouring ones. The new scenario of FPs, thus moderately blurred by the presence of non-zero tunnelling processes, is pictorially sketched in the lower row of figure \ref{fig:3_families} and fully discussed in \ref{sec:App_Stat_Config}.

In the following, we shall assume that the two condensates feature the same dynamical parameters, i.e. $U_a=U_b=:U$, $T_a=T_b=:T$ and $N=M$. Nevertheless, we note in advance that the presence of small deviation $T_a\neq T_b$, $U_a\neq U_b$ and $N\neq M$, from the previous ideal conditions (deviations which could be present in a real experimental setup), have been proved (by means of numerical simulations) not to significantly affect the developed analysis and the dynamical scenario discussed in the following. Both the hopping amplitudes and the interaction strengths depend on the lattice constants and on the scattering length \cite{Gerbier}, which are tunable parameters of the experimental setups mentioned in the introduction. In particular, interaction strengths can be controlled by means of Feshbach resonances \cite{RMP_Dalfovo, Yukalov}.

\section{Stability of stationary demixed states}
\label{sec:Stability_of}
Energetic-stability and linear-stability analysis are standard but powerful tools to investigate the qualitative dynamical behaviour of the system in the vicinity of a FP \cite{Arnold, Moser}. In view of an experimental realization, the developed analysis plays an important role, since it is impossible to prepare the system in a state which \textit{exactly} coincides to one of the aforementioned stationary configurations. Preliminary, it is convenient to introduce vector
$$
   \vec{z}=\left(\Phi_1,\,\Phi_2,\,\Phi_3,\Psi_1,\,\Psi_2,\,\Psi_3,\,n_1,\,n_2,\,n_3,\,m_1,\,m_2,\,m_3\right),
$$
and to write dynamical system (\ref{eq:Dyn_sys}) in the compact form 
\begin{equation}
   \dot{\vec{z}}= \mathbb{E} \vec{\nabla} \tilde{H}
   \label{eq:z_dot}
\end{equation}
where 
$$
\mathbb{E}= \left( \begin{array}{ccc}
 0_{6} & \mathbb{I}_6  \\
 -\mathbb{I}_6  &  0_{6} 
\end{array} \right)
$$
is the standard symplectic matrix,
\begin{equation}
    \tilde{H}=H-\lambda_a \sum_{j=1}^3n_j -\lambda_b \sum_{j=1}^3m_j
    \label{eq:tilde_H}
\end{equation}
is the effective Hamiltonian and $\vec{\nabla}\tilde{H}=(\partial_{\Phi_1}\tilde{H}\dots,\partial_{m_3}\tilde{H})$.

\subsection{Energetic stability}
\label{sec:Energetic_stability}
 An effective way to determine whether a FP $\vec{z}_*$ is energetically stable or not is to study the signature of the relevant Hessian matrix
\begin{equation}
    \label{eq:Hessian}
    \mathbb{H}_{i,j}(\vec{z}_*)=\left. \frac{\partial^2 \tilde{H}}{\partial z_i \partial z_j } \right|_{\vec{z}_*}.
\end{equation}
According to Lagrange-Dirichlet Theorem, a FP $\vec{z}_*$ is energetically stable if $\mathbb{H}(\vec{z}_*)$ is positive or negative definite \cite{Moser} (to be more precise, in the same spirit of \cite{Arwas}, one has to exclude the pair of vanishing eigenvalues corresponding to the two conserved quantities or, equivalently, consider a $8\times8$ Hessian matrix obtained after explicitly introducing constraints $\Phi_1=\Psi_1=0$, $n_1=N-n_2-n_3$ and $m_1=M-m_2-m_3$). A point exhibiting energetic instability is therefore neither a local minimum nor a local maximum of the energy function $\tilde{H}$ \cite{Arwas}. 
With reference to the first row of figure \ref{fig:All_plots_compressed}, one can observe that no FP of the class ``SDW - Soliton" is energetically stable, each of them being a multidimensional-saddle point for Hamiltonian function $\tilde{H}$ (see second panel). Interestingly, the energetically-stable region relevant to FPs of the class ``Dimer - Soliton" (see first panel) exactly corresponds to one of the three kinds of ground states that were found and discussed in \cite{NoiSREP}. In fact, all FPs $\vec{z}_*$ in such region (depicted in green) are indeed global minima of function $\tilde{H}$. Eventually, observing the third panel, one can recognize the presence of an energetically-stable region for moderately low values of $W/U$ (depicted in blue). In this region, FPs $\vec{z}_*$ are local maxima of function $\tilde{H}$. 
White regions in figure \ref{fig:All_plots_compressed} correspond to those values of $W/U$ and $T/(UN)$ for which FPs belonging to a given class do not exist. Such regions stand in between different sub-classes which differ in the relative phases $\Phi_j-\Phi_{j-1}$ between the wells. Each sub-class is a portion of parameters' space where stationary configurations share common features (e.g. the relative phases) and which are delimited by white regions. Fox example, the first class includes three sub-classes which, in turn, include points 1A, 1B and 1C respectively.

\subsection{Linear stability}
The linear stability (also called dynamical stability \cite{Arwas}) of a FP $\vec{z}_*$ of motion equations (\ref{eq:z_dot}), (i.e. a configuration such that $\dot{\vec{z}}_*=0$) is determined by the eigenvalues of Jacobian matrix \cite{Arnold}
\begin{equation}
    \label{eq:Jacobian}
    \mathbb{J}_{i,j}(\vec{z}_*) = \left. \frac{\partial \dot{z}_i}{\partial z_j}\right|_{\vec{z}_*}.
\end{equation}
More precisely, as dynamical system (\ref{eq:z_dot}) is a Hamiltonian one, a FP $\vec{z}_*$ is said to be linearly stable (or elliptic) if all eigenvalues of $\mathbb{J}_{i,j}(\vec{z}_*)$ are purely imaginary; conversely, it is said to be linearly unstable if at least one (pair of) eigenvalues of matrix (\ref{eq:Jacobian}) has non-zero real part. In the second row of figure \ref{fig:All_plots_compressed}, obtained by sweeping model parameters $W/U$ and $T/(UN)$, we have represented, for each of the three classes of stationary points characterized by demixing, the largest real part among the eigenvalues of matrix (\ref{eq:Jacobian}). 
 One can notice wide regions of the parameters' space where FPs of the class ``Dimer - Soliton" are linearly stable (represented in dark blue). Interestingly, while in the first sub-class (the one including point 1A), all FPs are linearly stable, in the remaining two sub-classes (respectively including points 1B and 1C) there are regions featuring linear stability and regions featuring linear instability. FPs of the class  ``SDW - Soliton" (see second panel), are mostly linearly unstable, excepts for a tiny triangular-like region existing only for $W/U>2$ (notice that FPs obtained in the unphysical situation $T=0$ are linearly stable too). As shown in the third panel, the vast majority of FPs belonging to the class ``Soliton - Soliton" are linearly stable, except for those ones in a narrow band confining with the white region and those ones in a needle-like region present for $W/U\approx2$ and moderately high values of $T/(UN)$.

\subsection{Scope of the energetic- and the linear-stability analysis}
\label{sec:scope}
If a trajectory moves away from a FP, the energetic- and the linear-stability analysis thereof are of little use. For this reason, one should employ other indicators such as the first Lyapunov exponent, which is the gold standard to distinguish regular and chaotic trajectories. A further limitation affecting the energetic- and the linear-stability analysis comes from their \textit{local} character. More specifically, also in view of an experimental realization, one should  pay particular attention to the \textit{size} of the FP's neighborhood where they are valid. Both aspects are discussed in section \ref{sec:Regular_and_chaotic}.

An important remark is in order concerning the traditional criterion to evaluate the linear stability of a FP (i.e. all eigenvalues $\lambda_j$'s of matrix (\ref{eq:Jacobian}) must be purely imaginary). Interestingly, the latter fails when the characteristic frequencies $\omega_j=\mathcal{I}\{\lambda_j\}$ satisfy a certain \textit{commensurability condition}, essentially represented by a diophantine equation (see \cite{Moser} for details, in particular for the procedure which is used to determine their sign). In this case, in fact, a so-called ``elliptic" FP ceases to be the center of an elliptic island and turns unstable. More specifically, it has been proven by Moser \cite{Moser} that, if the initial configuration $\vec{z}(t=0)=:\vec{z}_0$ is sufficiently close to a linearly stable FP $\vec{z}_*$, solutions of the actual non linear system (\ref{eq:z_dot}) \textit{almost always} depart from those of the linearized one only extremely slowly, if at all. Nevertheless, this is true only if characteristic frequencies $\omega_j$'s, properly taken with a certain sign, do not satisfy the aforementioned commensurability condition. The frequency vectors $\vec{\omega}$ which satisfy such condition constitute a dense set, although of measure zero, excepts in the positive ($\omega_j>0$ $\forall j$) and negative ($\omega_j<0$ $\forall j$) quadrants of space $\vec{\omega}$.
These two quadrants exactly corresponds to the regions where energetic stability holds.

\section{Regular and chaotic oscillations of boson populations}
\label{sec:Regular_and_chaotic}
For each of the 102729 pairs of model parameters $(W/U, T/(UN))$, a starting point $\vec{z}_0$ very close to the relevant FP $\vec{z}_*$ is chosen in such a way that the relative difference between the vector components of $\vec{z}_0$ and the corresponding ones of $\vec{z}_*$ is from $2\%$ to $5\%$ thus emulating what could be achieved in a real experimental set up \cite{Anker, Albiez, Milburn, Penna_PRL}. Then motion equations (\ref{eq:z_dot}) are numerically solved\footnote{Computational resources provided by HPC@POLITO (\texttt{http://www.hpc.polito.it})} for a series of consecutive time intervals and the first Lyapunov exponent is iteratively computed according to the standard scheme described in \cite{Sprott}. The comparison between the results (see third row of figure \ref{fig:All_plots_compressed}) and the previously discussed linear-stability analysis (see second row of figure \ref{fig:All_plots_compressed}) indeed shows that if a FP $\vec{z}_*$ is linearly unstable, than a trajectory starting from a point $\vec{z}_0$ close to it is chaotic, i.e. it is associated to a non-zero Lyapunov exponent. Note to the reader: in the following, labels 1A,...,3C are indistinctly used to indicate both a FP or a trajectory starting in a neighborhood thereof. 
Likewise, the vast majority of FPs featuring linear stability is such that the trajectory originating from a point $\vec{z}_0$ close to it is regular. Actually, for a limited number of FPs this is not true. For example, with reference to the central column of \ref{fig:All_plots_compressed}, the tiny linearly stable region present in the second row has no counterpart in the third row, all the trajectories therein represented being chaotic. This circumstance can be interpreted in terms of \textit{size} of the elliptic islands centered around an elliptic FP. As already evidenced in figure 7 of reference \cite{Penna_PRE}, such elliptic islands are very small when their center is a FP of the class ``SDW - Soliton" and so the distance $|\vec{z}_0-\vec{z}_*|$, despite chosen to be small, is already greater than the islands' characteristic radius. Moreover, it is worth noticing the presence of curved lines featuring a large Lyapunov exponent (e.g., in the first panel of the third row of figure \ref{fig:All_plots_compressed}, the curve whose bounds are points $(0.5,0.075)$ and $(2,0)$ in parameters' space $(W/U,T/(UN))$) which are expected to correspond to linearly stable FPs (see first panel of the second row). The seed of chaotic behavior is, in this case, the commensurability of characteristic frequencies $\omega_j$'s, which are the imaginary parts of the eigenvalues of Jacobian (\ref{eq:Jacobian}) (see \cite{Moser} for details, in particular for the procedure which is used to determine their sign). In fact, one can verify that all FPs constituting the aforementioned curve are such that $\omega_1=-2\omega_2$.

\section{How to quantify mixing and demixing of boson populations}
\label{sec:Quantify_mixing}
\begin{figure}[p!]
 \centering
 \includegraphics[width=\linewidth]{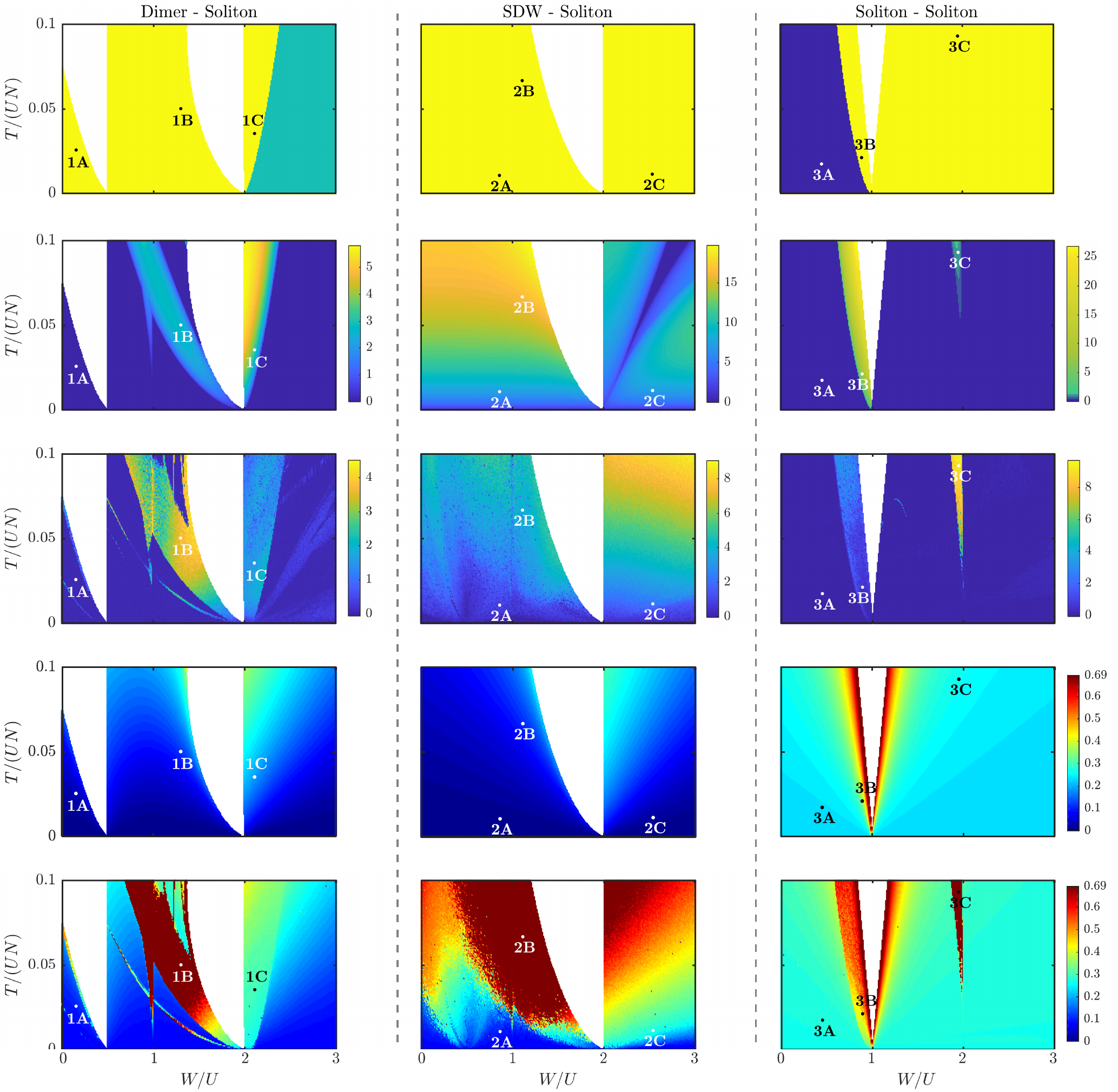}
\caption{\textbf{First row}: Energetic-stability analysis of FPs $\vec{z}_*$. Green and blue correspond to energetically stable regions, as they represent, respectively, minima and maxima of Hamiltonian (\ref{eq:tilde_H}). Yellow refers to energetically unstable regions, i.e. to saddle points of Hamiltonian (\ref{eq:tilde_H}). \textbf{Second row:} Linear-stability analysis of FPs $\vec{z}_*$. The color corresponds to $\max_{j}\{\mathcal{R}\{\lambda_j\}\}$ where $\lambda_j$'s are the eigenvalues of matrix (\ref{eq:Jacobian}). Dark blue represents linearly stable FPs. \textbf{Third row:} First Lyapunov exponent associated to trajectories starting close to FPs $\vec{z}_*$. Dark blue (yellow) corresponds to regular motions (highly chaotic trajectories). \textbf{Fourth row:} FPs' entropy of mixing, i.e. $S_{mix}(\vec{z}_*)$. Blue (red) represents a remarkable phase separation (an high degree of mixing). \textbf{Fifth row:} Maximum entropy of mixing, i.e. $\max_{0<t<50} \left\{S_{mix}(t)\right\}$ relevant to trajectories starting close to FPs $\vec{z}_*$. Blue corresponds to trajectories which feature a very small degree of mixing $S_{mix}$ throughout all the simulated dynamics. Conversely, red is associated to trajectories which, one or more times during the time evolution, feature complete mixing. \textbf{All panels:} White regions correspond to model parameters $W/U$ and $T/(UN)$ for which no stationary solutions of the type defined in the column's title exist. In each panel, three points have been highlighted in order to facilitate the discussion. Model parameters $N=50$ and $U=1$ have been chosen.}
 \label{fig:All_plots_compressed}
\end{figure}

Looking at the first row of figure \ref{fig:3_families}, one can recognize that the three presented configurations feature perfect demixing, as the presence of a condensed species in a certain well always implies the \textit{complete} absence of the other species. In the second row of the same figure, where the aforementioned ideal configurations are blurred by the activation of tunnelling processes, the two species, despite being still overall separated, feature a small degree of mixing. In fact, in whichever well a certain species is macroscopically present, the other one is \textit{nearly} absent, yet non zero. In the following we present an indicator to \textit{quantify} the degree of separation or, to be more precise, the degree of mixing. Entropy of mixing, generally denoted with $S_{mix}$, is a standard indicator commonly used in Statistical Thermodynamics when investigating miscibility properties of chemical compounds \cite{Adkins,Brandani,Camesasca} whose role, in the present work, is played by quantum gases. As the geometry behind the extended Bose-Hubbard model we are investigating is inherently discretized, one has to compute the entropy of mixing in each well and then evaluate the average over the three wells. One therefore obtains that 
\begin{equation}
    \label{eq:S_{mix}}
    S_{mix}=-\frac{1}{3}\sum_{j=1}^3\left[\frac{n_j}{n_j+m_j}\log\left(\frac{n_j}{n_j+m_j}\right) +  \frac{m_j}{n_j+m_j}\log\left(\frac{m_j}{n_j+m_j}\right) \right].
\end{equation}
From the definition itself of $S_{mix}$, two important properties, which serve to highlight the lower and the upper bounds of this indicator, emerge: i) The entropy of mixing of any perfectly demixed configuration, as the ones depicted in the upper row of figure \ref{fig:3_families}, is zero; ii) The entropy of mixing of a uniform configuration (i.e. any configuration $\vec{z}_{un}$ such that $n_j=m_j=N/3$, for $j=1,2,3$) features the maximum possible entropy of mixing, which reads
\begin{equation}
    \label{eq:S_uniform}
    S_{mix}(\vec{z}_{un})= -\frac{1}{3}\,6\,\frac{1}{2}\log\frac{1}{2}=\log 2 \approx 0.6931.
\end{equation}
In passing, we observe that the relative phases between the wells play no role in the computation of $S_{mix}$, but are crucial for its time evolution. The fourth row of figure \ref{fig:All_plots_compressed} shows the value of the entropy of mixing for all FPs belonging to the three classes discussed in section \ref{sec:Notable_classes}.
As expected, we observe that $S_{mix}$ steadily increases with $T/(UN)$, since a bigger hopping amplitude blurs the fully demixed configurations depicted in the upper row of figure \ref{fig:3_families}. Contrary to expectations, there are regions where $S_{mix}$ increases for increasing inter-species repulsion $W/U$ (to be more specific, the region featuring $1<W/U<2$ in the first panel, the region for $0<W/U<2$ in the second panel and the region featuring $0<W/U<1$ in the third panel). This circumstance looks counter intuitive, but it is easily explained by recalling that FPs are not necessarily minimum-energy configurations. As illustrated in the second row of figure \ref{fig:All_plots_compressed} and discussed in section \ref{sec:Energetic_stability}, in fact, a FP can also be a local maximum or a saddle point for effective Hamiltonian (\ref{eq:tilde_H}).  In the third panel, moreover, we observe that there are values of $W/U$ and $T/(UN)$ for which $S_{mix}$ tends to the maximum value $\log 2$. This happens because FPs of the class ``Soliton - Soliton" have been so blurred by the activation of the hopping amplitude $T$, that they have almost lost their identifying aspect and turned into a uniform configurations of the type $\vec{z}_{un}$ (as shown in figure \ref{fig:Fixed_points_Sol_Sol}, notice that the stationary configuration continuously varies with respect to model parameters $W/U$ and $T/(UN)$).

\section{Competition between phase separation and chaotic behaviour}
\label{sec:Despite_chaos}
In the same spirit of section \ref{sec:Regular_and_chaotic}, we have numerically solved motion equations (\ref{eq:z_dot}) choosing, for each parameters' pair $(W/U,T/(UN))$, a starting point $\vec{z}_0$ close to the corresponding FP $\vec{z}_*$. The choice has been made in such a way that the difference between the vector components of $\vec{z}_*$ and those of $\vec{z}_0$ is from $2\%$ to $5\%$. The knowledge of the time evolution of boson populations $n_j(t)$ and $m_j(t)$ allows one to readily compute $S_{mix}(t)$ and so to monitor the mixing properties of the atomic species.

If the initial configuration $\vec{z}_0$ lies within a regular island centered around a linearly stable FP $\vec{z}_*$ (and the characteristic frequencies thereof do not match Moser's commensurability condition \cite{Moser}), the motion consists in small oscillations around $\vec{z}_*$. Therefore, the entropy of mixing $S_{mix}(t)$ features small oscillations around the constant values $S_{mix}(\vec{z}_*)$ which, in turn, is very low for the vast majority of FPs $\vec{z}_*$ belonging to the three classes of notable stationary demixed configurations under considerations (recall the fourth row of figure \ref{fig:All_plots_compressed}). On the other hand, if $\vec{z}_0$ lies outside regular islands, the motion is chaotic, as discussed in section \ref{sec:Regular_and_chaotic}. In these circumstances, one would expect the two condensed species to fully mix and thus to quickly loose memory of their initial, demixed character. We show that this is not always the case, as demixing and chaos can coexist indeed. To this purpose, for each simulated trajectory, we have recorded $\max_{t} \left\{S_{mix}(t)\right\}$, where $t$ ranges from $t=0 s$ to $t=50 s$, a time interval whose width is three orders of magnitudes larger than the smallest characteristic period of populations' oscillations. Overall, 102729 trajectories have been simulated, each one starting from an initial condition $\vec{z}_0$ which, in turn, is close to a FP $\vec{z}_*$. The result is shown in the fifth row of figure \ref{fig:All_plots_compressed}. 

\begin{figure}[htb!]
 \centering
 \includegraphics[width=\linewidth]{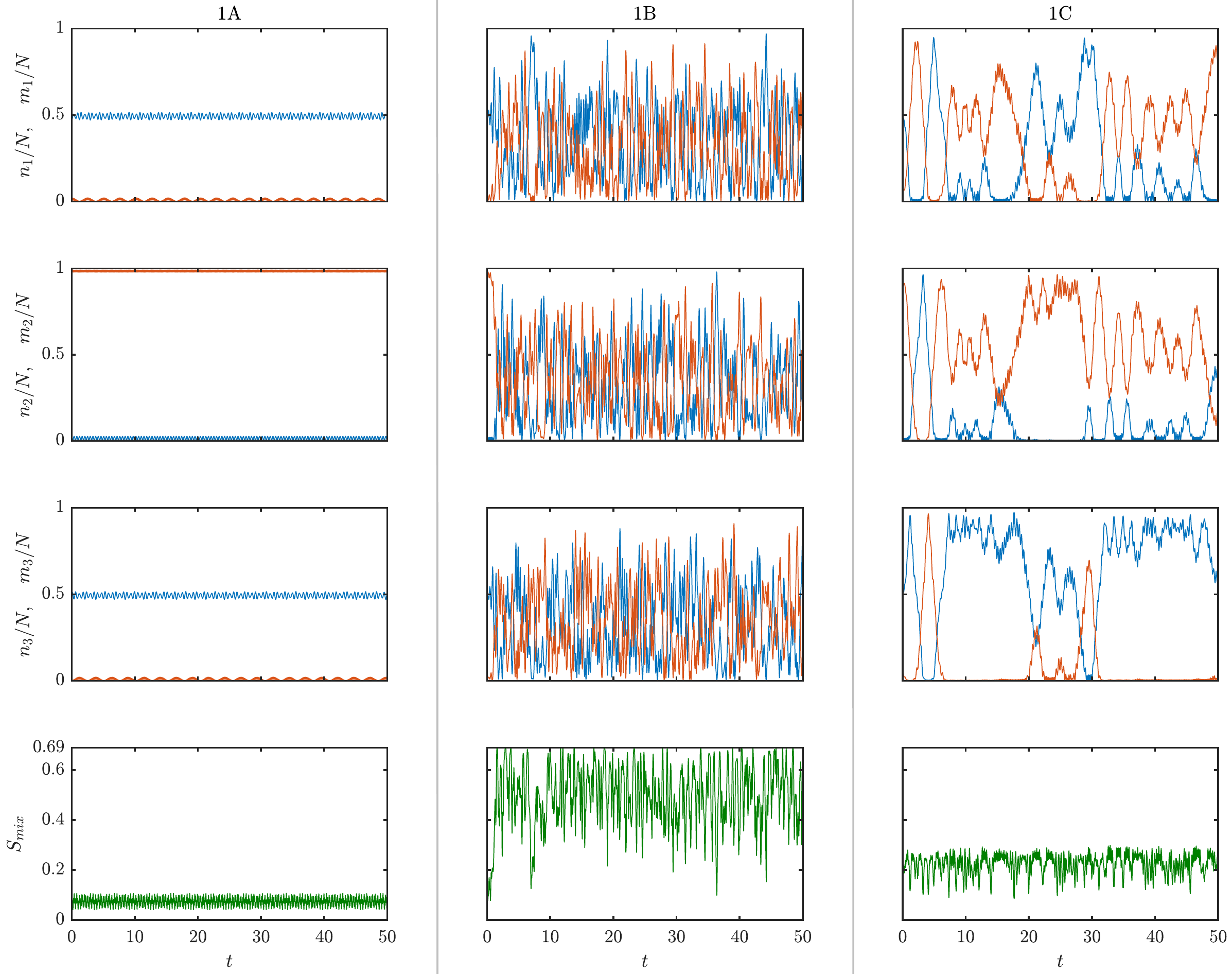}
 \caption{Time evolution of normalized boson populations and of entropy of mixing $S_{mix}$. The results have been obtained numerically solving equations \ref{eq:z_dot}. Blue (red) denote species-A (B) bosons. Each column corresponds to the dynamics originating from three different starting points $\vec{z}_0$ which, in turn, are chosen in the vicinity of FPs 1A, 1B and 1C respectively.   }
 \label{fig:Dynamic_Dim_Sol}
\end{figure}

\begin{figure}[htb!]
 \centering
 \includegraphics[width=\linewidth]{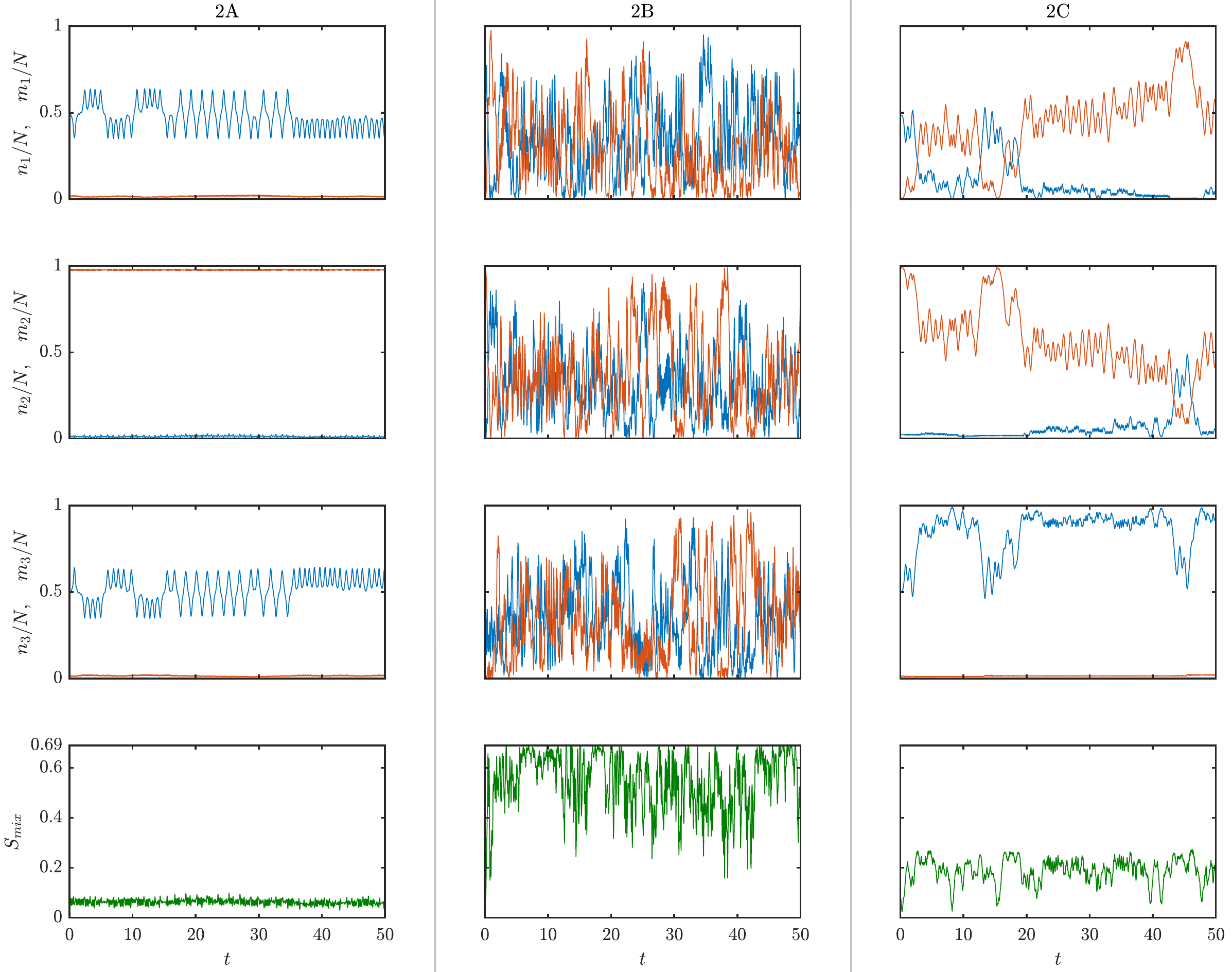}
 \caption{Time evolution of normalized boson populations and of entropy of mixing $S_{mix}$. The results have been obtained numerically solving equations \ref{eq:z_dot}. Blue (red) denote species-A (B) bosons. Each column corresponds to the dynamics originating from three different starting points $\vec{z}_0$ which, in turn, are chosen in the vicinity of FPs 2A, 2B and 2C respectively.   }
 \label{fig:Dynamic_SDW_Sol}
\end{figure}

\begin{figure}[htb!]
 \centering
 \includegraphics[width=\linewidth]{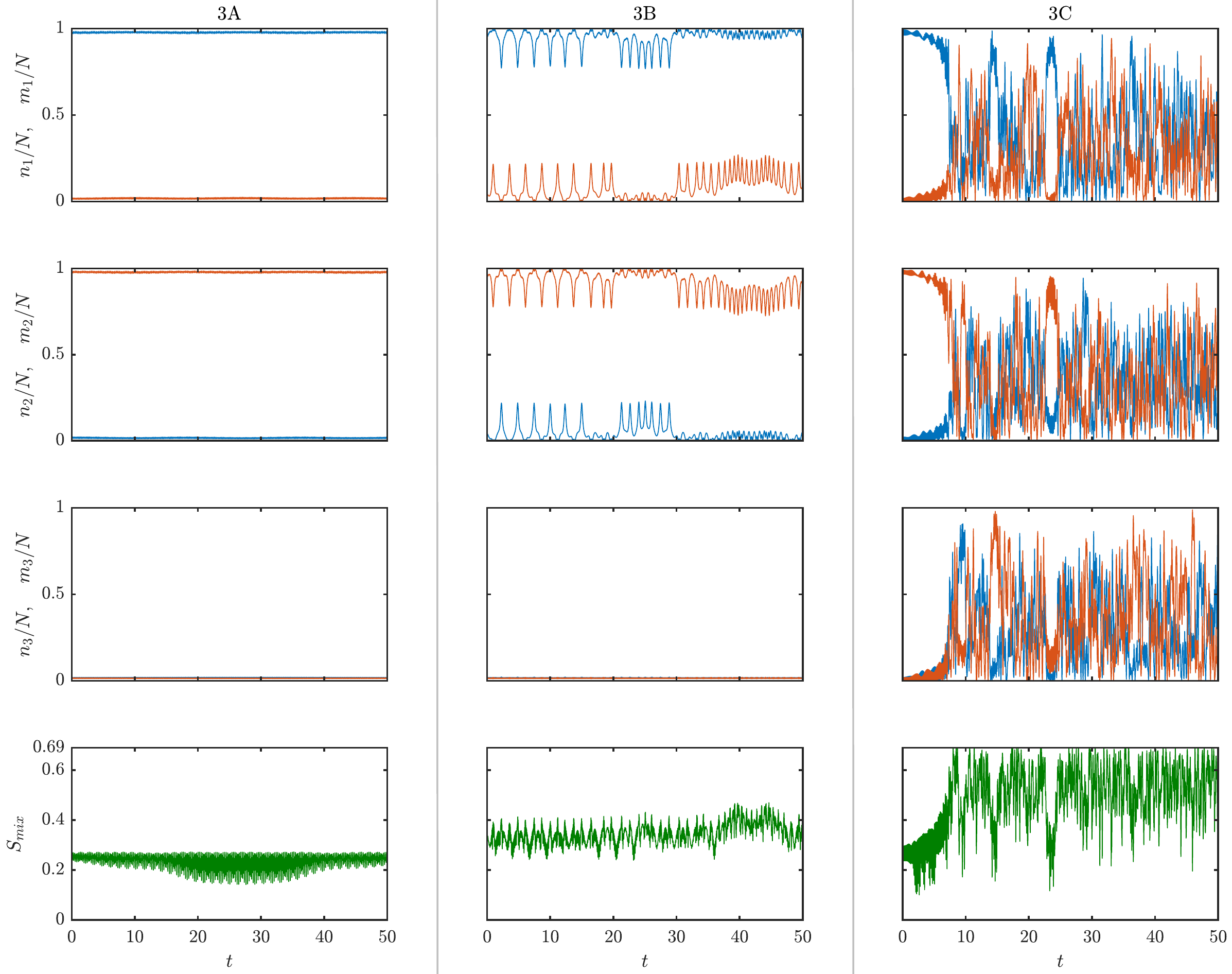}
 \caption{Time evolution of normalized boson populations and of entropy of mixing $S_{mix}$. The results have been obtained numerically solving equations \ref{eq:z_dot}. Blue (red) denote species-A (B) bosons. Each column corresponds to the dynamics originating from three different starting points $\vec{z}_0$ which, in turn, are chosen in the vicinity of FPs 3A, 3B and 3C respectively.   }
 \label{fig:Dynamic_Sol_Sol}
\end{figure}

In the attempt to facilitate the comparison among the info provided by the linear-stability analysis (second row of figure \ref{fig:All_plots_compressed}), by the computation of the first Lyapunov exponent (third row), by the evaluation of $S_{mix}(\vec{z}_*)$ (fourth row) and of $\max_t\{S_{mix}(t)\}$ (fifth row), we highlight and analyze some notable dynamical regimes and we explicitly illustrate them in figures \ref{fig:Dynamic_Dim_Sol}-\ref{fig:Dynamic_Sol_Sol}. We group them according to the regularity of the motion and to the persistence of demixing during the dynamics. We remind that labels 1A,...,3C are indistinctly used to indicate either a FP or a trajectory starting in a neighborhood thereof. 
\begin{itemize}
    \item \textit{Regular oscillations of demixed species.} The regimes represented in the first column of figure \ref{fig:Dynamic_Dim_Sol} and in the first column of figure \ref{fig:Dynamic_Sol_Sol} are regular, i.e. they feature a vanishing Lyapunov exponent. The initial states ($t=0$) lie in elliptic islands centered around linearly stable FPs $1A$ and $3A$, respectively (and their characteristic frequencies do not match Moser's commensurability condition \cite{Moser}). The time evolution of boson populations consists in small oscillations around FPs. As a consequence, $S_{mix}(t)$ slightly oscillates around constant values $S_{mix}(1A)\approx0.08$ and $S_{mix}(3A)\approx0.25$ respectively, thus witnessing the persistence of a remarkable demixing.

    \item \textit{Fully developed mixing.} The regimes illustrated in the second column of figure \ref{fig:Dynamic_Dim_Sol}, second column of figure \ref{fig:Dynamic_SDW_Sol} and third column of figure \ref{fig:Dynamic_Sol_Sol} are chaotic, i.e. they are associated to a non-zero Lyapunov exponent. The initial configurations lies in the vicinity of linearly unstable FPs. The onset of chaos completely destroys the original, demixed configurations whose entropies of mixing are approximately equal to the ones relevant to the corresponding FPs, i.e. $S_{mix}(1B)\approx 0.15$, $S_{mix}(2B)\approx 0.13$ and $S_{mix}(3C)\approx 0.23$ respectively. As a consequence, in all three cases, $S_{mix}(t)$ repeatedly reaches the maximum possible value of $\approx 0.69$ which, in turn, witnesses the full mixing of the bosonic species.

    \item \textit{Persistent demixing despite chaos.} The dynamical regimes depicted in the first column of figure \ref{fig:Dynamic_SDW_Sol} and in the second column of figure \ref{fig:Dynamic_Sol_Sol} consist in small chaotic oscillations around the FPs 2A and 3B respectively. Chaos develops because the initial conditions already lie in the chaotic sea which, in turn, has been shown to surround linearly unstable FPs. However, despite the occurrence of chaos, not only the demixing of bosonic species persists, but also the macroscopic structure of the initial configurations remains unchanged during the dynamics. As a consequence, the oscillations of $S_{mix}(t)$ are chaotic but their amplitude is small, namely it never exceeds critical values $\approx 0.10$ and $\approx 0.44$ respectively. 
    
    The dynamical regimes illustrated in the third column of figure \ref{fig:Dynamic_Dim_Sol} and in the third column of figure \ref{fig:Dynamic_SDW_Sol} are chaotic as well, but much less shrunken. It is a fact that, also in these cases, despite the presence of chaos, the two atomic species feature a low degree of mixing for all the simulated dynamics (in fact $S_{mix}(t)$ remains smaller than $\approx 0.29$ and $\approx 0.27$ respectively). Nevertheless, contrary to trajectories 2A and 3B (which consist in small chaotic oscillations around the equilibrium points), in these cases, chaos disrupts the structure of the original configurations, repeatedly triggering \textit{populations inversions}. In other words, for what concerns trajectories 1C and 2C, as time goes by, boson populations in each well severely change but always in such a way to preserve the (low) degree of mixing.

\end{itemize}

\noindent
Table \ref{tab:summary} is intended to summarize the aforementioned results.
\begin{table}[h!]
\centering
\label{tab:summary}
\begin{tabular}{|c|c|c|c|c|c|c|}
\hline
 & \begin{tabular}[c]{@{}c@{}}Energetic\\ stability\end{tabular} &  \begin{tabular}[c]{@{}c@{}}Linear\\ stability\end{tabular} &
 \begin{tabular}[c]{@{}c@{}}Chaotic/ \\ regular\end{tabular} & 
 \begin{tabular}[c]{@{}c@{}}Initial\\ demixing \end{tabular}  &  \begin{tabular}[c]{@{}c@{}}Persistent\\ demixing\end{tabular} &  \begin{tabular}[c]{@{}c@{}}Energetically\\ inaccessible mixing \end{tabular}    \\ \hline
 1A & \xmark  & \cmark & Regular  & \cmark & \cmark & \xmark \\ \hline
 1B & \xmark  & \xmark & Chaotic  & \cmark & \xmark & \xmark \\ \hline
 1C & \xmark  & \xmark & Chaotic  & \cmark & \cmark & \cmark \\ \hline
 2A & \xmark  & \xmark & Chaotic  & \cmark & \cmark & \xmark \\ \hline
 2B & \xmark  & \xmark & Chaotic  & \cmark & \xmark & \xmark \\ \hline
 2C & \xmark  & \xmark & Chaotic  & \cmark & \cmark & \cmark \\ \hline
 3A & \cmark  & \cmark & Regular  & \cmark & \cmark & \cmark \\ \hline
 3B & \xmark  & \xmark & Chaotic  & \cmark & \cmark & \xmark \\ \hline
 3C & \xmark  & \xmark & Chaotic  & \cmark & \xmark & \xmark \\ \hline
\end{tabular}
\caption{Summary of the most important static and dynamical features for each of the nine representative cases.}
\end{table}

Interestingly, we observe that the persistence of spatial phase separation marks extended bundles of chaotic trajectories, i.e. chaotic trajectories involving wide patches of starting points and rather extended ranges of model parameters. Such a persistence can be explained either in terms of energy conservation or by recalling the presence of regular islands where chaotic trajectories cannot enter.

\begin{figure}[htb!]
 \centering
 \includegraphics[width=\linewidth]{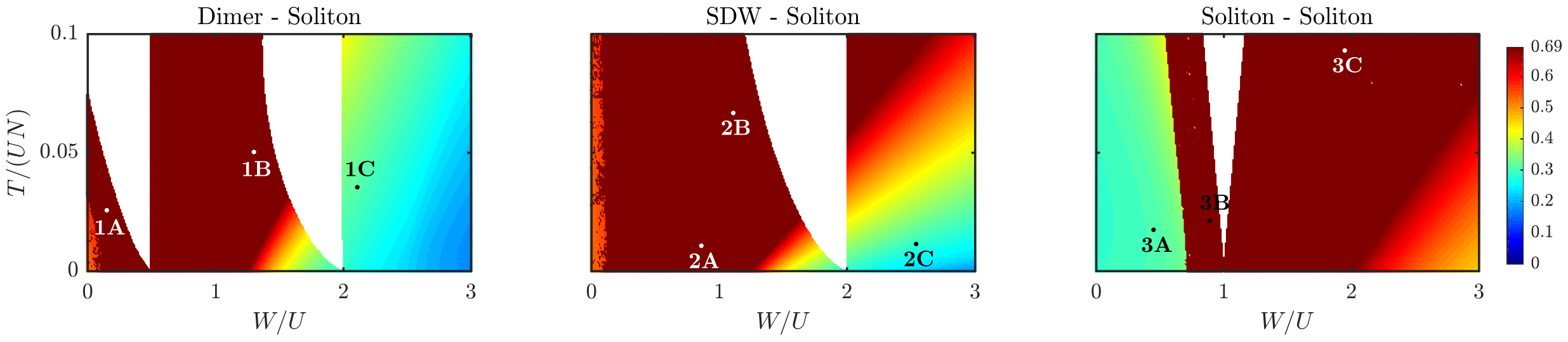}
 \caption{Maximum entropy of mixing $\bar{S}_{mix}$ over constant-energy hypersurfaces $\Gamma$'s which, in turn, are well specified by choosing an initial condition $\vec{z}_0$ and a  pair of model parameters $(W/U,T/(UN))$. The maximum value of the entropy of mixing over an hypersurface $\Gamma$, $\bar{S}_{mix}$, is computed according to the standard method of Lagrange multipliers, a technique which allows to find the maximum of a certain objective function in presence of one or more constraints. White regions correspond to model parameters $W/U$ and $T/(UN)$ for which no stationary solutions of the type defined in the column's title exist. In each panel, three points have been highlighted in order to facilitate the discussion. Model parameters $N=50$ and $U=1$ have been chosen.}
 \label{fig:Max_S_Ergo}
\end{figure}

Concerning the energy-conservation argument, we remind that the choice of a certain $\vec{z}_0$ automatically fixes the trajectory and the constant-energy hypersurface $\Gamma$ where the trajectory will be confined. Then, we analytically determine the maximum value of the entropy of mixing over the entire $\Gamma$, $\bar{S}_{mix}$, and note that the entropy of mixing along the trajectory, $S_{mix}(t)$, will never exceed $\bar{S}_{mix}$ at any time. In other words, (see also \ref{sec:App_Phase_space}, where we comment on the structure of phase space), the trajectory will wander wide regions of $\Gamma$ but, if $\bar{S}_{mix}$ is small enough, it will never visit highly-mixed configurations. The values $\bar{S}_{mix}$ are determined over surfaces $\Gamma$ by maximizing the objective function $S_{mix}$ under the constraints $\sum_{j}n_j=N$, $\sum_{j}m_j=N$ and $H=H(\vec{z}_0)$, for each initial state $\vec{z}_0$. Such computation of $\bar{S}_{mix}$ as a function of model parameters is based on the well-known method of Lagrange multipliers. The result is shown figure \ref{fig:Max_S_Ergo}.
As an example, compare the green domain around point 1C in figure \ref{fig:Max_S_Ergo} (witnessing persistent spatial phase phase separation) with the light-blue domain around point 1C in the third row of figure \ref{fig:All_plots_compressed} (displaying a manifestly chaotic behaviour). The same reasoning holds also for point 2C. 

Concerning the regular-islands argument, the latter is suggested by several examples of trajectories which, despite exhibiting a chaotic behaviour together with persistent spatial phase separation, are associated to the maximum possible (i.e. energetically accessible) value of $\bar{S}_{mix}:=\max_{\Gamma}\left\{S_{mix}\right\}$. For example, the trajectory 3B is of this type since, as shown in figure \ref{fig:Max_S_Ergo}, the constant-energy hypersurface $\Gamma$ where it is embedded, features the biggest possible entropy of mixing, $\log 2$ (depicted in dark red). A reasonable interpretation of this apparent mismatch is linked to the possible presence of regular islands on $\Gamma$ \cite{Kandrup} (see also \ref{sec:App_Phase_space}). If $\bar{S}_{mix}$ lies inside such islands, in fact, chaotic trajectories will never have the chance to visit highly-mixed configurations.
This interpretation can be applied also to point 2A. A detailed analysis to ascertain the presence of mixed configurations inside regular islands requires an extended work that will be developed elsewhere.

\section{Concluding remarks}
\label{sec:Conclusions}
We have investigated the dynamics of a bosonic binary mixture loaded in a three-well potential with periodic boundary conditions, its relation with the entropy of mixing and the robustness of spatial phase separation. In general, the developed analysis, even if focused on some particular classes of configurations, has provided a considerable amount of information about dynamical regimes characterized by regular and chaotic behaviours. 

In section \ref{sec:A_binary_mixture}, we have introduced the model describing the mixture in the ring trimer and derived the corresponding semiclassical motion equations. Section \ref{sec:Notable_classes} has been devoted to the presentation of the three notable classes of stationary configurations featuring demixing, ``Dimer - Soliton", ``SDW - Soliton" and ``Soliton - Soliton".  In section \ref{sec:Stability_of}, we have developed the energetic- and the linear-stability analysis of the previously identified stationary configurations, highlighting their scope and limitations. In section \ref{sec:Regular_and_chaotic}, we have explicitly computed the first Lyapunov exponent along 102729 trajectories starting in the vicinity of as many FPs thus clearly identifying regular and chaotic regimes. We have observed that chaos can originate in three different ways: 1) When the trajectory starts in the vicinity of a linearly unstable FP; 2) When the trajectory starts in the neighbourhood of an elliptic FP such that its characteristic frequencies match Moser's commensurability condition \cite{Moser}; 3) When the initial configuration lies outside the regular island centered around a linearly stable FP.  In section \ref{sec:Quantify_mixing}, we have introduced the entropy of mixing $S_{mix}$, borrowed from Statistical Thermodynamics \cite{Camesasca}, to quantify the degree of mixing between the two condensed species. 

Eventually, in section \ref{sec:Despite_chaos} we have shown that the chaotic motion of boson populations and demixing can coexist or, in other words, that chaos, despite present, may not be able to completely disrupt the order imposed by phase separation. Such coexistence can occur either because highly mixed states lie in regular islands where the chaotic trajectory cannot penetrate or because the constant-energy hypersurface does not contain mixed states at all.  In conclusion, we notice that our study could be of interest both for further theoretical investigations and for future experiments with bosonic mixtures trapped in ring lattices which, as discussed in the introduction, are within the reach of current experimental technology.

\appendix

\section{Stationary configurations featuring demixing}
\label{sec:App_Stat_Config}

\begin{figure}[htb!]
 \centering
 \includegraphics[width=\linewidth]{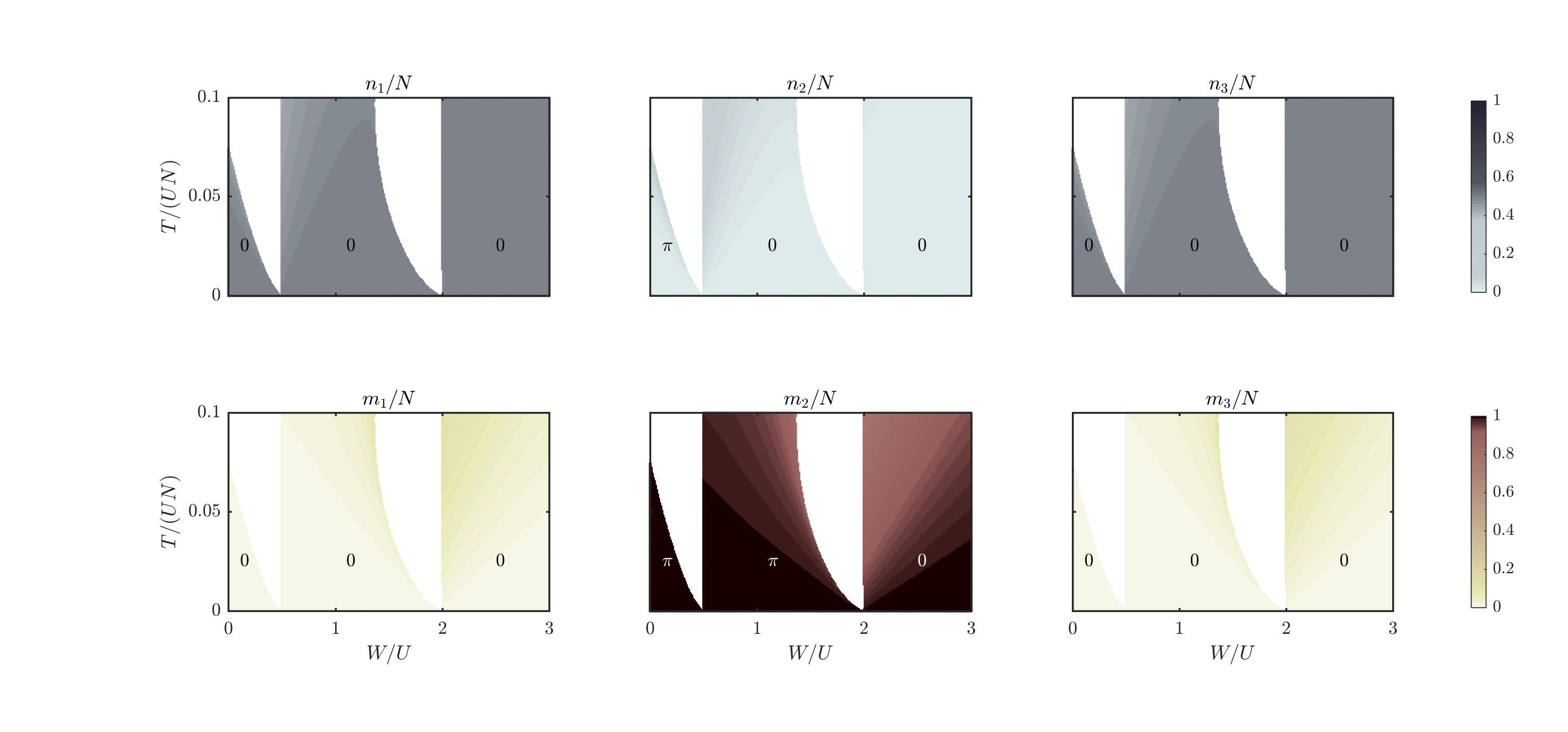}
 \caption{Class of FPs of the type Dimer - Soliton. Each column corresponds to a well while each row to a different condensed species. The color corresponds to the fraction of bosons hosted by the well (see color bars) while numbers $0$ and $\pi$ indicate the phase shift with respect to the first well. In the white regions, this kind of stationary configuration does not exist. }
 \label{fig:Fixed_points_Dim_Sol}
\end{figure}
\begin{figure}[htb!]
 \centering
 \includegraphics[width=\linewidth]{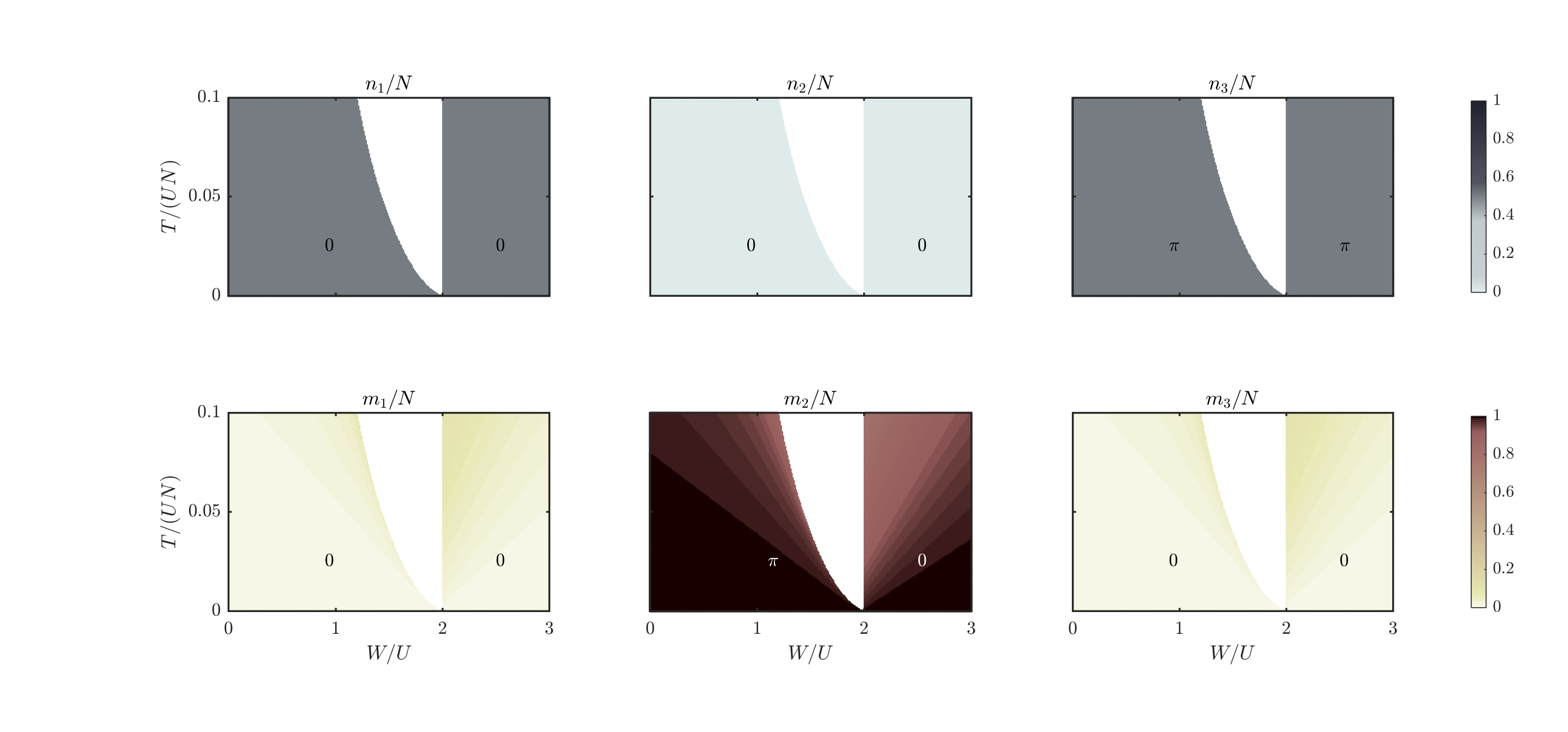}
 \caption{Class of FPs of the type SDW - Soliton.  Each column corresponds to a well while each row to a different condensed species. The color corresponds to the fraction of bosons hosted by the well (see color bars) while numbers $0$ and $\pi$ indicate the phase shift with respect to the first well. In the white region, this kind of stationary configuration does not exist. }
 \label{fig:Fixed_points_SDW_Sol}
\end{figure}
\begin{figure}[htb!]
 \centering
 \includegraphics[width=\linewidth]{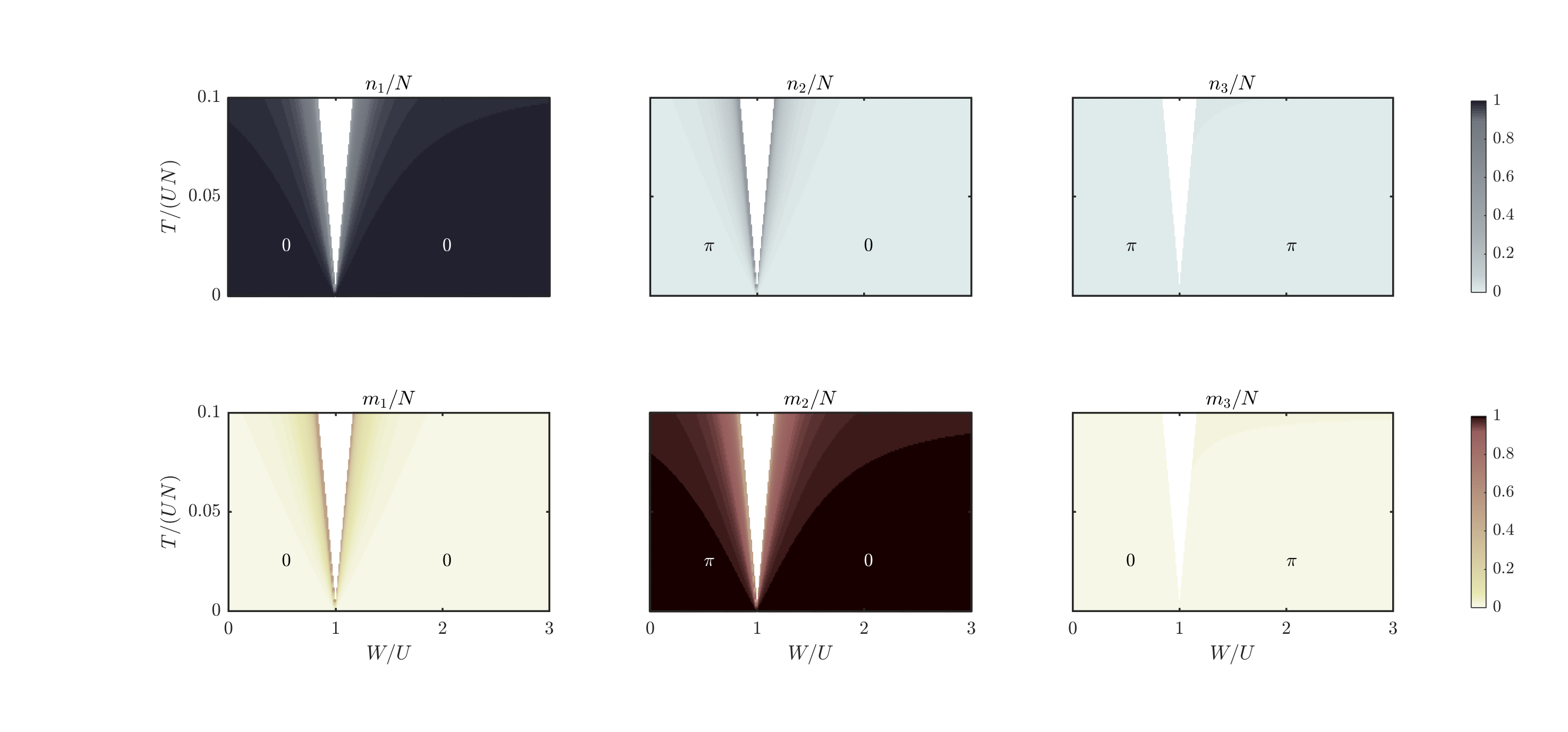}
 \caption{Class of FPs of the type Soliton - Soliton.  Each column corresponds to a well while each row to a different condensed species. The color corresponds to the fraction of bosons hosted by the well (see color bars) while numbers $0$ and $\pi$ indicate the phase shift with respect to the first well. In the white region, this kind of stationary configuration does not exist. }
 \label{fig:Fixed_points_Sol_Sol}
\end{figure}

Figures \ref{fig:Fixed_points_Dim_Sol}-\ref{fig:Fixed_points_Sol_Sol} show in detail how the stationary configuration changes in a non-narrow range of model parameters. They account both for the variations of boson populations $n_j$, $m_j$ (different color shades) and for the relative phase between the wells (numbers within the various regions). Notice that white areas, representing the absence of a certain class of FPs, stand in between different sub-classes. The latter differ in the relative phases between the wells and are three in the ``Dimer - Soliton" case, two in the ``SDW - Soliton" case and in the ``Soliton - Soliton" case as well (see figures \ref{fig:Dynamic_Dim_Sol}, \ref{fig:Dynamic_SDW_Sol} and \ref{fig:Dynamic_Sol_Sol} respectively). We conclude by observing that the collective phase difference between the two condensed species play no role in the dynamics of the system (see Hamiltonian (\ref{eq:Hamiltoniana_classica})), so figures \ref{fig:Dynamic_Dim_Sol}-\ref{fig:Dynamic_Sol_Sol} have been drawn arbitrarily setting $\Phi_1=\Psi_1=0$.

\section{Remarks on the dynamical behaviour of trajectories starting close to a fixed point}
\label{sec:App_qualitative_behaviour}
In the following, we describe the qualitative behaviour of a trajectory which, at $t=0$, starts from a point $\vec{z}_0$ very close to a FP $\vec{z}_*$. To this end we consider 4 different kinds of FPs $\vec{z}^*$ corresponding to different kinds of stability/instability.

\paragraph{Linearly-stable FP.}
The solutions of the linearized system of differential equations $\dot{\vec{y}}=\mathbb{J}(\vec{z}_*)\,\vec{y}$ induced by matrix (\ref{eq:Jacobian}) correspond to small oscillations around a given equilibrium point $\vec{z}_*$  \cite{Arnold}, the characteristic frequencies thereof being $\omega_j=\mathcal{I}\{\lambda_j\}$, where $\lambda_j$'s are the eigenvalues of matrix (\ref{eq:Jacobian}). The effectiveness of the solutions of the linearized equations in representing those of the actual non-linear equations has been discussed in section \ref{sec:scope}.

\paragraph{Energetically-stable FP.}
Energetic stability is stronger than linear stability, as one can prove that if the initial configuration $\vec{z}_0$ is sufficiently close to FP $\vec{z}_*$, solutions of the actual non linear system (\ref{eq:z_dot}) remain arbitrarily close to those of the linearized one \textit{for all times} and, moreover, there are no issues associated to the commensurability of characteristic frequencies $\omega_j$'s.

\paragraph{Linearly-unstable FP.}
A point of linear instability (also called dynamic instability \cite{Arwas}) is such that almost every trajectory will depart from it. For a generic (i.e. non necessarily Hamiltonian) dynamical system, a trajectory starting close to an unstable FP can have any sort of behaviour (e.g. fall towards a FP, towards a periodic orbit, become chaotic, etc.). Since we are dealing with an \textit{Hamiltonian} system, one knows a priori that the relevant flow in the phase space is incompressible, so the number of possible alternatives decreases. Despite not rigorously proven \footnote{Consider, as a counter example, a particle in a one-dimensional double-well-like potential centered at $x=0$. Of course the local maximum present at $x=0$ is a linearly unstable FP. Nevertheless trajectories starting in a neighborhood thereof are regular.}, as the number of effective degrees freedom is relatively high ($D=4$), one can expect that linearly unstable FPs are surrounded by chaos (see \ref{sec:App_Phase_space}). The validity of this reasonable ansatz is confirmed by the explicit calculation of the first Lyapunov exponent along trajectories starting close to FPs $\vec{z}_*$ (compare second and third row of figure \ref{fig:All_plots_compressed}).

\paragraph{Energetically-unstable FP.} An energetically-unstable FP can be linearly stable (but not viceversa), so the qualitative behaviour of a trajectory starting from its neighborhood strongly depends on the linear stability of $\vec{z}_*$ and, as already discussed, on the possible commensurability of characteristic frequencies $\omega_j$'s. Although we have restricted our analysis to isolated systems at zero temperature, it is worth mentioning that, if dissipation is introduced or in presence of thermalization processes, an energetically-unstable system will tend to decay to an energetically-stable state \cite{Paraoanu}.

\section{On the structure of phase space}
\label{sec:App_Phase_space}
The phase space associated to Hamiltonian dynamical system \ref{eq:z_dot} seems to be 12-dimensional as, for each of the two condensed species and for each of the three wells, there are two canonically conjugate dynamic variables: local boson number $n_j$ ($m_j$) and local phase $\Phi_j$ ($\Psi_j$). As discussed in section \ref{sec:Notable_classes}, the relative phase between the two condensed species play no role in the dynamics (see Hamiltonian (\ref{eq:Hamiltoniana_classica})) so one can arbitrarily fix $\Phi_1=\Psi_1=0$. Moreover, as the total number of bosons $N=\sum_{j=1}^3n_j$ and $M=\sum_{j=1}^3m_j$ constitute two  independent conserved quantities, one can substitute $n_1=N-n_2-n_3$ and $m_1=M-m_2-m_3$. Therefore, the number of effective dynamical variables is $8$, which correspond to $D=4$ degrees of freedom. It is well known, also in the field of ultracold atoms \cite{Arwas}, that there is a profound difference between systems featuring $D=2$ and $D>2$ degrees of freedom \cite{Lieberman}. Only in the first case, the three-dimensional space corresponding to the constant-energy ($\tilde{H}=E$) hypersurface can be divided by the relevant two-dimensional KAM tori into separated regions. Chaotic trajectories, if present, are therefore always topologically confined by KAM tori. For $D>2$, instead, the latter cannot divide the phase space into separated regions (in the same way as a circumference cannot divide the euclidean space $\mathbb{R}^3$ into two parts). All chaotic regions are therefore interconnected by a very slow percolation-like phenomenon which goes under the name of Arnold diffusion \cite{Lieberman, Arwas}.
Nevertheless, they do not occupy the whole constant-energy space, since regular islands are still present (e.g. the neighbourhoods of energetically-stable FPs), the relative measure thereof being still an open problem \cite{Kandrup}. In passing, we mention that, when the number of degrees of freedom of a non linear dynamical system tends to infinite, the measure of regular islands tends to zero and so the constant-energy hypersurface gets completely chaotic, thus justifying the ergodic hypothesis and a microcanonical approach to the problem \cite{Kandrup}.

\section*{References}
\providecommand{\newblock}{}


\begin{thebibliography}{100}
\expandafter\ifx\csname url\endcsname\relax
  \def\url#1{{\tt #1}}\fi
\expandafter\ifx\csname urlprefix\endcsname\relax\def\urlprefix{URL }\fi
\providecommand{\eprint}[2][]{\url{#2}}

\bibitem{cmixt3}
\"Ohberg P and Stenholm S 1998 {\em Phys. Rev. A\/} {\bf 57}(2) 1272--1279

\bibitem{cmixt4}
Esry B~D and Greene C~H 1999 {\em Phys. Rev. A\/} {\bf 59}(2) 1457--1460

\bibitem{cmixt5}
Svidzinsky A~A and Chui S~T 2003 {\em Phys. Rev. A\/} {\bf 67}(5) 053608

\bibitem{kasa}
Kasamatsu K and Tsubota M 2006 {\em Phys. Rev. A\/} {\bf 74}(1) 013617

\bibitem{Gallemi_1}
Melé-Messeguer M, Julia-Diaz B, Guilleumas M, Polls A and Sanpera A 2011 {\em
  New Journal of Physics\/} {\bf 13} 033012

\bibitem{tic}
Ticknor C 2013 {\em Phys. Rev. A\/} {\bf 88}(1) 013623

\bibitem{lee}
Lee K~L, J\o{}rgensen N~B, Liu I~K, Wacker L, Arlt J~J and Proukakis N~P 2016
  {\em Phys. Rev. A\/} {\bf 94}(1) 013602

\bibitem{Inguscio}
Thalhammer G, Barontini G, De~Sarlo L, Catani J, Minardi F and Inguscio M 2008
  {\em Phys. Rev. Lett.\/} {\bf 100}(21) 210402

\bibitem{Gadway}
Gadway B, Pertot D, Reimann R and Schneble D 2010 {\em Phys. Rev. Lett.\/} {\bf
  105}(4) 045303

\bibitem{Soltan}
Soltan-Panahi P, L{\"u}hmann D~S, Struck J, Windpassinger P and Sengstock K
  2012 {\em Nature Physics\/} {\bf 8} 71

\bibitem{jz}
Jaksch D and Zoller P 2005 {\em Annals of Physics\/} {\bf 315} 52 -- 79 ISSN
  0003-4916 special Issue

\bibitem{bdz}
Bloch I, Dalibard J and Zwerger W 2008 {\em Rev. Mod. Phys.\/} {\bf 80}(3)
  885--964

\bibitem{yuk}
Yukalov V~I 2009 {\em Laser Physics\/} {\bf 19} 1--110

\bibitem{sep2}
Jain P and Boninsegni M 2011 {\em Phys. Rev. A\/} {\bf 83}(2) 023602

\bibitem{sep3}
Lingua F, Guglielmino M, Penna V and Capogrosso~Sansone B 2015 {\em Phys. Rev.
  A\/} {\bf 92}(5) 053610

\bibitem{Angom}
Suthar K and Angom D 2016 {\em Phys. Rev. A\/} {\bf 93}(6) 063608

\bibitem{NoiPRA2}
Penna V and Richaud A 2017 {\em Phys. Rev. A\/} {\bf 96}(5) 053631

\bibitem{ks}
Kuklov A~B and Svistunov B~V 2003 {\em Phys. Rev. Lett.\/} {\bf 90}(10) 100401

\bibitem{ddl}
Duan L~M, Demler E and Lukin M~D 2003 {\em Phys. Rev. Lett.\/} {\bf 91}(9)
  090402

\bibitem{qe1}
Roscilde T and Cirac J~I 2007 {\em Phys. Rev. Lett.\/} {\bf 98}(19) 190402

\bibitem{pol}
Benjamin D and Demler E 2014 {\em Phys. Rev. A\/} {\bf 89}(3) 033615

\bibitem{Tempere}
Casteels W, Tempere J and Devreese J~T 2013 {\em Phys. Rev. A\/} {\bf 88}(1)
  013613

\bibitem{ent}
Wang W, Penna V and Capogrosso-Sansone B 2016 {\em New Journal of Physics\/}
  {\bf 18} 063002

\bibitem{NoiEntropy}
Lingua F, Richaud A and Penna V 2018 {\em Entropy\/} {\bf 20} 84

\bibitem{Penna_NJP}
Lingua F, Lepori L, Minardi F, Penna V and Salasnich L 2018 {\em New Journal of
  Physics\/} {\bf 20} 045001

\bibitem{Catani}
Catani J, Lamporesi G, Naik D, Gring M, Inguscio M, Minardi F, Kantian A and
  Giamarchi T 2012 {\em Phys. Rev. A\/} {\bf 85}(2) 023623

\bibitem{EPL}
Johnson T~H, Bruderer M, Cai Y, Clark S~R, Bao W and Jaksch D 2012 {\em EPL
  (Europhysics Letters)\/} {\bf 98} 26001

\bibitem{Makarov}
Makarov D and Uleysky M 2017 {\em Communications in Nonlinear Science and
  Numerical Simulation\/} {\bf 43} 227 -- 238 ISSN 1007-5704

\bibitem{modul}
Jin G~R, Kim C~K and Nahm K 2005 {\em Phys. Rev. A\/} {\bf 72}(4) 045601

\bibitem{Catani_deg}
Catani J, De~Sarlo L, Barontini G, Minardi F and Inguscio M 2008 {\em Phys.
  Rev. A\/} {\bf 77}(1) 011603

\bibitem{Soltan_2}
Soltan-Panahi P, Struck J, Hauke P, Bick A, Plenkers W, Meineke G, Becker C,
  Windpassinger P, Lewenstein M and Sengstock K 2011 {\em Nature Physics\/}
  {\bf 7} 434

\bibitem{Amico}
Amico L, Osterloh A and Cataliotti F 2005 {\em Phys. Rev. Lett.\/} {\bf 95}(6)
  063201

\bibitem{Agamalian}
Aghamalyan D, Amico L and Kwek L~C 2013 {\em Phys. Rev. A\/} {\bf 88}(6) 063627

\bibitem{Anker}
Anker T, Albiez M, Gati R, Hunsmann S, Eiermann B, Trombettoni A and Oberthaler
  M~K 2005 {\em Phys. Rev. Lett.\/} {\bf 94}(2) 020403

\bibitem{Albiez}
Albiez M, Gati R, F\"olling J, Hunsmann S, Cristiani M and Oberthaler M~K 2005
  {\em Phys. Rev. Lett.\/} {\bf 95}(1) 010402

\bibitem{PennaLinguaJPB}
Lingua F, Mazzarella G and Penna V 2016 {\em Journal of Physics B: Atomic,
  Molecular and Optical Physics\/} {\bf 49} 205005

\bibitem{PennaLinguaPRE}
Lingua F and Penna V 2017 {\em Phys. Rev. E\/} {\bf 95}(6) 062142

\bibitem{NoiSREP}
Penna V and Richaud A 2018 {\em Scientific Reports\/} {\bf 8} 10242

\bibitem{Penna_Amico}
Amico L and Penna V 1998 {\em Phys. Rev. Lett.\/} {\bf 80}(10) 2189--2192

\bibitem{Nonlinearity}
Franzosi R, Livi R, Oppo G~L and Politi A 2011 {\em Nonlinearity\/} {\bf 24}
  R89

\bibitem{Eilbeck}
Eilbeck J, Lomdahl P and Scott A 1985 {\em Physica D: Nonlinear Phenomena\/}
  {\bf 16} 318 -- 338 ISSN 0167-2789

\bibitem{Penna_PRE}
Franzosi R and Penna V 2003 {\em Phys. Rev. E\/} {\bf 67}(4) 046227

\bibitem{Johansson}
Jason P, Johansson M and Kirr K 2012 {\em Phys. Rev. E\/} {\bf 86}(1) 016214

\bibitem{Olsen}
Olsen M~K and Corney J~F 2016 {\em Phys. Rev. A\/} {\bf 94}(3) 033605

\bibitem{Arwas}
Arwas G, Vardi A and Cohen D 2015 {\em Scientific Reports\/} {\bf 5} 13433

\bibitem{Kolovsky}
Kolovsky A~R 2016 {\em International Journal of Modern Physics B\/} {\bf 30}
  1630009

\bibitem{Ellinas}
Ellinas D, Johansson M and Christiansen P~L 1999 {\em Physica D: Nonlinear
  Phenomena\/} {\bf 134} 126 -- 143 ISSN 0167-2789

\bibitem{Kevrekidis}
Kevrekidis P~G 2009 {\em The discrete nonlinear Schr{\"o}dinger equation:
  mathematical analysis, numerical computations and physical perspectives\/}
  vol 232 (Springer Science \& Business Media)

\bibitem{Gerbier}
Gerbier F, Widera A, F\"olling S, Mandel O, Gericke T and Bloch I 2005 {\em
  Phys. Rev. A\/} {\bf 72}(5) 053606

\bibitem{RMP_Dalfovo}
Dalfovo F, Giorgini S, Pitaevskii L~P and Stringari S 1999 {\em Rev. Mod.
  Phys.\/} {\bf 71}(3) 463--512

\bibitem{Yukalov}
Yukalov V~I 2004 {\em Laser Physics Letters\/} {\bf 1} 435

\bibitem{Arnold}
Arnol'd V~I 2013 {\em Mathematical methods of classical mechanics\/} vol~60
  (Springer Science \& Business Media)

\bibitem{Moser}
Moser J 1958 {\em Communications on Pure and Applied Mathematics\/} {\bf 11}
  81--114

\bibitem{Milburn}
Milburn G~J, Corney J, Wright E~M and Walls D~F 1997 {\em Phys. Rev. A\/} {\bf
  55}(6) 4318--4324

\bibitem{Penna_PRL}
Buonsante P, Franzosi R and Penna V 2003 {\em Phys. Rev. Lett.\/} {\bf 90}(5)
  050404

\bibitem{Sprott}
Sprott J~C 2003 {\em Chaos and time-series analysis\/} vol~69 (Citeseer)

\bibitem{Adkins}
Adkins C~J 1983 {\em Equilibrium thermodynamics\/} (Cambridge University Press)

\bibitem{Brandani}
Brandani G~B, Schor M, MacPhee C~E, Grubm{\"u}ller H, Zachariae U and
  Marenduzzo D 2013 {\em PloS one\/} {\bf 8} e65617

\bibitem{Camesasca}
Camesasca M, Kaufman M and Manas-Zloczower I 2006 {\em Macromolecular theory
  and simulations\/} {\bf 15} 595--607

\bibitem{Kandrup}
Kandrup H~E, Sideris I~V and Bohn C~L 2001 {\em Phys. Rev. E\/} {\bf 65}(1)
  016214

\bibitem{Paraoanu}
Paraoanu G~S 2003 {\em Phys. Rev. A\/} {\bf 67}(2) 023607

\bibitem{Lieberman}
Lichtenberg A~J and Lieberman M~A 2013 {\em Regular and stochastic motion\/}
  vol~38 (Springer Science \& Business Media)

\end{thebibliography}
\end{document}